\documentstyle[prl,eqsecnum,epsf,aps,psfig]{revtex} 
\newcommand{\bc}{\begin{center}}
\newcommand{\ec}{\begin{center}}
\newcommand{\eq}{\begin{equation}}
\newcommand{\en}{\end{equation}}

\newcommand{\xe}{$^{136}$Xe }

\begin{document}

\tighten
\draft
\twocolumn %take this out before submission

\title{Search for Strange Matter by Heavy Ion Activation}

\author{M.C. Perillo Isaac, Y.D. Chan, R. Clark, M.A. Deleplanque,
M.R. Dragowsky$^{*}$, P. Fallon, I.D. Goldman$^{\dagger}$,
R-M. Larimer, I.Y. Lee, A.O. Macchiavelli,
R.W. MacLeod, K. Nishiizumi$^{\ddag}$,
E.B. Norman, L.S. Schroeder,
F.S. Stephens}

\address{Lawrence Berkeley National Laboratory - Berkeley CA 94720\\
$^*$Oregon State University, Corvallis OR 97331 \\
$^{\dagger}$University of Sao Paulo, Sao Paulo, Brazil \\
$^{\ddag}$Space Sciences Laboratory, University of California, Berkeley CA 94720
}
\maketitle
\begin{abstract}
We present the results of 
an experimental search for stable 
strange matter 
using the heavy ion activation technique.
We studied samples of a meteorite,
terrestrial nickel ore, and lunar soil. 
Our search improved the existing experimental
limit on the strange matter content in normal matter 
by 2 to 3 orders of
magnitude, and allowed us to probe for the first time 
the flux of low mass strangelets on the
lunar surface.
\end{abstract}

Suggestions of various forms of tightly bound
strongly-interacting matter have been made in the past\cite{ho76}.
Strange matter, aggregates of up, down, and strange 
quarks, are a theoretically possible form of these systems.
If strange matter exists and is absolutely stable,
it would be the true ground state of the strong interaction.
E. Witten\cite{ew84}
raised the possibility that particles of 
stable strange matter, also called strangelets,
would be a significant dark matter candidate. 
De R\'ujula and Glashow\cite{rg84}
suggested different methods 
to detect strangelets in the Earth and in
space-based
experiments.
Later, Alcock and
Farhi\cite{af85}
placed severe restrictions on scenarios 
for strange matter survival in
the hot temperatures of the early universe.
Nevertheless, there is no evidence against
the possibility that strange matter is stable and,
although not produced cosmologically, 
is present in today's Universe. It should, therefore, 
be possible to probe its concentrations
in Earth-based experiments.

A favorable astrophysical environment for the
formation of strange matter would be inside neutron
stars\cite{gle95}\cite{alp86}.
In fact, if strange matter is stable, all neutron stars are "strange
stars"\cite{al88}. 
The decay of the orbits of binary pairs of such compact
stars lead to their collision, allowing for a fraction
of their material to be injected into the galaxy\cite{gle90}\cite{bo93}.

Experimental searches for strange
matter have been performed using a variety of techniques,
sensitive to different strangelet mass ranges. 
Searches were performed in cosmic ray experiments, where the
strangelets would show anomalous energy 
loss in matter\cite{lo91}\cite{bu88}\cite{liu}.
Experimental limits on the concentration of 
strange matter in normal matter are
due to Br\"ugger and 
collaborators\cite{br89}, for strangelets in the
$400<$A$<10^{7}$~amu mass range.
Accelerator mass spectroscopy 
is the most sensitive technique for detecting 
low mass strangelets (A$<300$), 
relying on the assumption that 
strangelets would possess the same chemical properties
as normal nuclei with the same charge, behaving as ultra-heavy
isotopes \cite{van96}.
It was also pointed out that 
low mass strangelets, produced in
heavy ion collisions, would be evidence of the production of
quark and gluon plasma\cite{bar91}. More speculatively, 
they could be "grown" by neutron
absorption and could be used as an energy 
source\cite{go89}, since strange matter absorbs normal
matter exothermically. 

The properties of stable strange matter were calculated by 
Farhi and Jaffe
\cite{fj85a}. Berger and Jaffe\cite{be87} developed a mass formula 
for strangelets and studied stable configurations and
possible decay modes of highly excited
strangelets.
Extensions to this model were performed by Takahasi and Boyd
\cite{ta88}. Stable strange matter has positive,
but lower charge than ordinary matter for the same mass.
Consequently it presents a lower Coulomb
barrier than ordinary nuclear matter,
leading Farhi and Jaffe \cite{fj85} to propose 
a method to search for strange matter via heavy ion
activation.
In such an experiment, when normal matter penetrates the 
Coulomb barrier of a strangelet, 
the quarks in normal matter will "dissolve" inside
the strangelet, releasing energy. 
The energy added to the system is
given by $\Delta E = IA_{B} + K $, 
where $I$ is the extra binding energy per
nucleon of strange matter relative to that 
of normal matter, $A_{B}$ is the mass of the beam nucleus,
and $K$ is the kinetic energy of the beam. $I$ could be as large as
5 to 20~MeV\cite{jpc}, meaning that energies of the order of GeV's
can be released in the interaction. Nevertheless, some of 
this energy, $E_{M}$, will be used by the system
to regain flavor equilibrium \cite{be87}.
The remaining energy available will be released in the form of photons. 

The argument against the emission of nucleons from excited
strangelets can be understood as follows:
particle emission requires that the deconfined quarks 
inside the strangelet gain the configuration of a particle, say a neutron.
This would imply an improbably high local concentration of energy, 
and as shown in \cite{be87} is an unimportant decay mode for strangelets
with $A>2000$. 
Similarly, the mechanism for pion production, through the creation of 
a quark-antiquark pair, requires a very energetic quark near the boundary
of the strangelet \cite{ban83}. 
Particle emission, such as observed in heavy-ion, collision are
also unlikely, since subthreshold pion production and pre-equilibrium nucleon
emission have very low cross sections, specially for low energy 
projectiles \cite{shu}\cite{ver}.

The excited strangelet is modeled by a fermi gas
with uniform 
temperature $T=\left( (2\mu \delta E) /(\pi^{2} A) \right)^{1/2}$,
characteristic of the photon spectrum emitted
in the de-excitation.
In this equation, $A$ is the baryon number of the strangelet, $\mu$ is 
the quark chemical potential (roughly 300 MeV),
and $\delta E = \Delta E - E_{M}$, with $\Delta E$ and $E_{M}$ previously
defined. According to \cite{fj85}, the strangelets are not opaque to photons 
with energies characteristics of these temperatures, implying that 
the spectrum of photons emitted will be similar but not equal to 
the spectrum of a cooling black body. 
Depending on its mass, the excited strangelet will radiate many low
energy photons, indicating that such an experiment requires a detector 
with a large solid angle,
high granularity, and sensitivity
to a broad energy range.

The GAMMASPHERE\cite{gsref} detector array
is an ideal instrument to perform this search.
GAMMASPHERE is a gamma-ray detector 
array composed of 110 elements.
Each element has a high-purity germanium (Ge) detector surrounded by
bismuth germanate (BGO) crystals.
The $4\pi$ solid angle coverage is shared by the Ge
detector (45\%) 
and the BGO crystals (55\%).
The Ge detectors are sensitive to a wide range of energies, 
from 20~keV to 20~MeV, 
while the BGO crystals cover the energy range from 20~keV to 10~MeV.
Furthermore, our sensitivity to low
energy photons, of the order of 20~keV, is 
enhanced in a strange matter interaction by the
high multiplicity of low energy photons, generating 
high energy deposition in the detectors due to pile-up.

We performed our experiment at the 88-Inch Cyclotron, 
using \xe at 450~MeV, delivered at 
250~enA. Three samples from distinct origins were examined: 
nickel ore found at 2070~m
underground\cite{ore}, the Allende meteorite\cite{met},
and lunar soil collected in the Apollo-17 mission\cite{m1}.
The lunar soil sample is composed of very fine grains,
and 200~mg of this soil was
compressed into an aluminum cup to produce a suitable target. 

Table 1 summarizes the data used to obtain our results.
The beam current could not be monitored continuously, since the beam
stopped in the thick insulating targets,
but periodic measurements of the beam current
performed upstream during the irradiation time confirmed the beam stability.
The range of the beam in the samples was 
calculated using TRIM\cite{trim}. 
The composition of all the samples is very similar, SiO$_{2}$
being
the main component.
The calculated ranges of the beam in these targets
are all of the order of 36$\mu$m.

The sensitivity of this experiment was evaluated by
Monte Carlo
using GEANT~3.21\cite{geref}.
The response of GAMMASPHERE to a strange 
matter signal was evaluated for $I =5$ MeV. 
In this case, if the interacting beam is $^{136}$Xe, 
the energy available, $\Delta E$, is 1.13~GeV.
$E_{M}$ was evaluated as a function of the strangelet mass
through the mass formula derived in \cite{be87}.
$E_{M}$ depends on the charge ($Z$) and 
hypercharge ($Y$) of the stable strangelet configuration. If the 
strangelet is large, the addition of a \xe nucleus should not change
its equilibrium flavor. 
We assumed that $\left| Z_{m}^{'} - Z_{m}\right| \ll 54$ and 
that $\left| Y_{m}^{'} - Y_{m}\right| \ll 136$, where $Z_{m}$ 
and $Y_{m}$ are the 
charge and the hypercharge of the stable strangelet before the 
addition of a \xe nucleus, and $Z_{m}^{'}$ and $Y_{m}^{'}$ are the charge
and hypercharge of the stable strangelet with $A^{'}=A+136$.
This approximation is valid for strangelets with masses $A\geq 2000$~amu.
According to ref.\cite{be87}, $Z_{m}=105$ for a strangelet of mass 
2000, and $Z_{m}^{'}=110$ for a strangelet of mass 2136. These values were 
obtained assuming
150~MeV for the mass of the strange quark, $m_{s}$ and 300~MeV 
for the up-quark chemical potential, $\mu_0$.

For the simulation, we also assumed that the spectrum 
of photons emitted is that of a black body characterised by 
the temperature $T$. Table 2 shows the net total energy available, $\delta E$, 
the characteristic strangelet temperature, and the 
expected number of photons released.

The characteristics of a strange matter signal are
high multiplicity and high energy deposition in 
the detector. We used four parameters to select strange matter event
candidates: the Ge detector multiplicity, NGE; 
the BGO detector multiplicity, NBGO;
the total energy deposited in all Ge detectors, $\Sigma$EGE; and the
total energy deposited in all BGO detectors, $\Sigma$EBGO.
These quantities are all functions of the total energy released in the
interaction and the individual energy of the photons released.
Thus, for the same beam impinging onto a strangelet,
they are functions of the baryon number of the strangelet and
the total energy available $\delta E$.

Figure 1 shows the comparison between the experimental 
distribution of NGE and $\Sigma$EGE 
and the distributions predicted by Monte
Carlo calculations. The experimental distributions correspond to the
bombardment of the lunar soil sample.
The distributions of NGE, NBGO, $\Sigma$EGE, and
$\Sigma$EBGO were obtained by the generation of 
100 events for strangelet masses ranging from 2000 to 
$10^{8}$ amu. These distributions were fitted to gaussian curves
and events from our data set were
selected if NGE, NBGO, $\Sigma$EGE, and
$\Sigma$EBGO are within two standard deviations from the
fitted Monte Carlo predictions. 
No events in our data satisfy these cuts, 
allowing us to set upper upper limits
on the concentration of strangelets in our samples.

We tested the efficiency for extraction of high multiplicity events in the
data using simulated events randomly inserted in the data set.
The extraction efficiency of these events
from the data set, using the event selection described above, is 100\%.
Pulser data was also acquired and analyzed at different frequencies and
amplitudes in order to verify the readout of high multiplicity events.

The concentration of strangelet in our samples, $n$,
is given by the
relation:
\eq
n=\frac{N}{\sigma \times r_{beam} \times p}
\en
\noindent
where $N$ is the number of events observed, $\sigma= \sigma_{0} A^{2/3}$ 
is the cross section for the interaction, 
$r_{beam}$ is the range of the
beam particles in the samples, and $p$ is the number 
of particles impinging the sample. $\sigma$ is purely geometric, and 
$\sigma_{0} = 3.04 \times 10^{-26}$ cm$^{2}$
is obtained assuming a baryon number density of 
0.25~fm$^3$\cite{af85}\cite{fj85}. 

In figure 2 we plot the limits obtained, assuming 1 as an 
upper limit for $N$,
for the three samples analyzed.
For comparison, the limits obtained
by Br\"ugger and collaborators in ref.\cite{br89} are also in the plot.
Our experiment was able to improve Br\"ugger's limit for
strangelet masses between $2000$ and $A=10^{7}$ by 3 orders
of magnitude, and to set limits for strangelets masses of 
the order of $A=10^{8}$, a range inaccessible
to Br\"ugger's experiment.
Our experiment was mostly sensitive to light strangelets, with masses
below $A=10^{9}$, which, if present as cosmic rays, would be
absorbed in the Earth's atmosphere.
Since the Moon has no atmosphere and its surface has been
exposed for millions of years, the upper limit in concentration 
of strange matter in the lunar soil allows
us to derive a limit for the flux of strangelets
impinging the surface of the Moon.

The sample used was collected at a depth of
upper 0.5 to 1 cm at
the base of the Sculptured Hills, Station 8\cite{m1}. Details on the analysis 
of this sample can be found in\cite{m2}\cite{m3}.
The presence of
high cosmic ray track densities in the sample 
indicates that the integrated
lunar surface exposure age is of the order of 100 My\cite{m4}. 
Taking into
account the range of
strange matter in normal matter suggested by De R\'ujula
and Glashow\cite{rg84},
the range of strange matter masses to which our experiment is sensitive,
and the integrated lunar surface exposure age, we are able to
estimate a limit on the flux of strangelets on the surface of the Moon.

Figure 3 shows the upper limits on the flux of strangelet
as a function of their
mass obtained by this experiment.
For comparison, the limits in the 
flux of strangelets obtained by Shirk and Price\cite{bp}.
Limits on the incoming flux of strangelets
in the Earth's atmosphere
obtained by Porter and collaborators
\cite{p85} are also consistent with those obtained by \cite{bp}.
Porter's limits were derived from four experiments originally carried out
to detect high energy 
cosmic gamma-rays, and are limited only to high-mass
strangelets.

The limit on the flux of strangelets in the surface of
the Moon allows us to set upper limits for the mass density of 
strangelets in the Galaxy.
If strangelets have a typical galactic velocity
of $3 \times 10^{7}$~cm/s, the upper limits for the
mass density of strangelets of different masses will 
vary from $3\times 10^{-37}$ g/cm$^3$  
for A=5000 strangelets, to $2\times 10^{-31}$ g/cm$^3$ for 
A=$10^{8}$ strangelets. These limits should be compared with the 
upper bound estimated by Glendenning based on the
collapse of binary compact stars\cite{gle90}, 
$10^{-29}$ g/cm$^3$. 
We note that that our results do not rule out the possibility
of the existence of strange matter in the Universe.
Even though our upper limits are 2 to 8 orders of magnitude lower than
the previous estimates, many quantities carry large
uncertainties, such as the fraction of pulsars that occur in binary 
compact systems and the fraction of mass ejected in binary stars 
collisions. 

\section*{Acknowledgements}

The authors would like to thank
B. Fujikawa for discussions about the data analysis,
R. Jaffe for his encouragement on the
preparation phase of the experiment,
C. Lyneis for cyclotron  discretionary time,
G. L. Shaw and N. Glendenning for their support and 
time for discussions, and A. Lyon for preparing
our lunar soil target.
This work was supported in part by the U.S. Department of Energy
under contract DEAC03-76SF00098,
and the National Aeronautics and Space Administration.

%%%%%%%%  REFERENCES

%%%%%%%%%%%%%%%%  TABLES

\begin{table}
\caption{Summary of data used in the analysis.
The \xe charge state in all irradiations was $26^{+}$. $<A>$ represents
the average beam current during irradiation}
\begin{tabular}{ccl}
Time  & $<A>$ & Target \\
(hours) & (nA) & \\
4.0  & 250 & Allende met.\\
5.1  & 250 & Ni ore \\
15.1 & 0   & BKG \\
13.2 & 220& Lunar soil \\
16.3 & 0  & BKG \\
\end{tabular}
\end{table}

\begin{table}
\caption{Characteristic signal expected from interactions of \xe
and strangelets of different masses. $\delta E$ is the energy released in the
form of photons, $T$ is the characteristic temperature of the photon
spectrum and $N_{\gamma}$ is the expected number of photons
emitted per strange matter event.}
\begin{tabular}{cccc}
$A$              & $\delta E$ & $T$ & $N_{\gamma}$ \\
(amu)            & (GeV)& (keV)  & \\
$2\times 10^{3}$ & 0.11 & 1855.3 & 61  \\
$5\times 10^{3}$ & 0.72 & 2961.3 & 144  \\
$1\times 10^{4}$ & 0.92 & 2370.0 & 391  \\
$1\times 10^{5}$ & 1.11 & 820.0  & 1353  \\
$1\times 10^{6}$ & 1.13 & 261.5  & 4314  \\
$1\times 10^{7}$ & 1.13 & 82.7   & 13653  \\
$1\times 10^{8}$ & 1.13 & 26.2   & 43178  \\
\end{tabular}
\end{table}

%%%%%%%%%%   FIGURES
\begin{figure}
%\begin{center}
\epsfxsize=0.44\textwidth
\epsffile{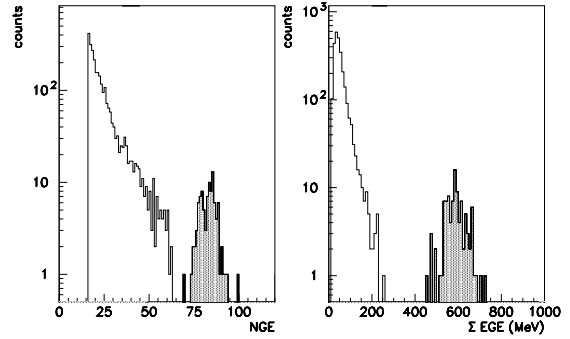}
\caption{Comparison between the measured distribution of NGE
and $\Sigma$EGE and the expected distributions for 
strange matter events. The simulated values (hatched) 
shown here we obtained for
100 interactions of \xe and strangelets of mass $10^{4}$ amu. 
The experimental distributions correspond to the bombardment 
of the lunar soil sample and require NGE$>15$ and NBGO$>15$.}
%\end{center}
\end{figure}

\begin{figure}
%\begin{center}
%\leavevmode
\epsfxsize=0.44\textwidth
\epsffile{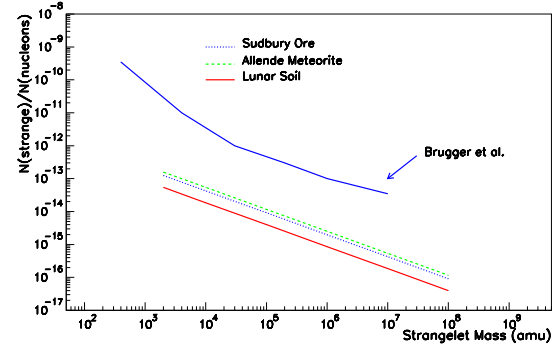}
\caption{Experimental limit on the concentration of strangelet in our samples.
The limits are based on the number of events which survive the
cuts described in the text, i.e., have a NGE, NBGO, $\Sigma$EGE and
$\Sigma$EBGO within 2 standard deviations from the expected values. 
The results from Br\"ugger
and collaborators obtained in an iron meteorite are also plotted for
comparison. $N_{strange}/N_{nucleons}$ is the concentration of strangelets
per nucleons in the sample.}
%\end{center}
\end{figure}

\begin{figure}
%\begin{center}
%\leavevmode
\epsfxsize=0.44\textwidth
\epsffile{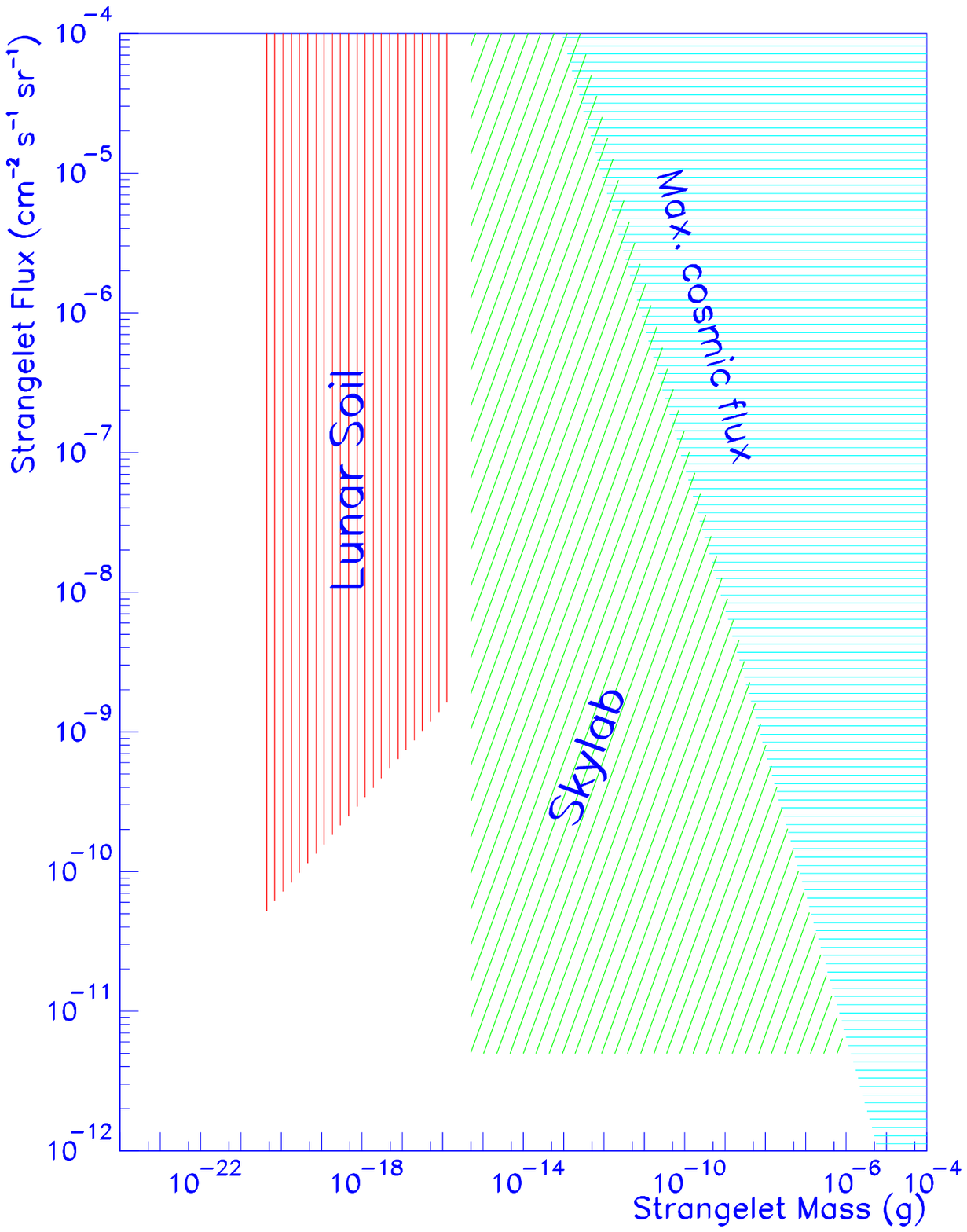}
\caption{Limits on the flux of strangelet impinging on the lunar surface
obtained by this experiment, and by Shirk and Price from a Lexan
array on the Skylab space station.Excluded regions are the hatched areas.
Maximum cosmic flux refers to the cosmic flux of strangelets assuming
that all the dark matter in the universe is
composed of strangelets.
We used the estimation of the range of strangelet given by De R\'ujula and
Glashow as a function of the strangelet mass to evaluate their range,
and considered the integrated
exposure of the lunar soil sample, $10^{8}$~years.}

%\end{center}
\end{figure}


\begin{references}

\bibitem{ho76} R.J. Holt, et al., Phys. Rev. Lett. {\bf 36}, 183, 1976, and
references therein.

\bibitem{ew84}E. Witten, Phys. Rev. D {\bf 30}, 272, 1984.

\bibitem{rg84}A. De R\'ujula and S.L. Glashow, Nature {\bf 312}, 734, 1984.

\bibitem{af85}C. Alcock and E. Farhi, Phys Rev. D {\bf 32}, 1273, 1985.

\bibitem{gle95} N. Glendenning, et al., 
Phys. Rev. Lett. {\bf 74}, 3519, 1995.

\bibitem{alp86} C. Alcock, et al.,
Phys. Rev. 
Lett. {\bf 57}, 2088, 1986.

\bibitem{al88} C. Alcock, and A. Olinto, Ann. Rev. Part. Sci. 
{\bf 38}, 161, 1988, and references therein.

\bibitem{gle90} N.K. Glendenning, Mod. Phys. Lett. A{\bf 5}, 2197, 1990.

\bibitem{bo93} R.N. Boyd and T. Saito, Phys. Lett. B{\bf 298}, 6, 1993.

\bibitem{lo91} D. Lowder, Nucl. Phys. B (Proc. Suppl.) {\bf 24B}, 84, 1991.

\bibitem{bu88} P.B. Price, Phys. Rev. D, {\bf 38}, 38, 1988.

\bibitem{liu} G. Liu and B. Barish, Phys. Rev. Lett. {\bf 61}, 271, 1988. 

\bibitem{br89}M. Br\"ugger et al., Nature {\bf 337}, 434, 1989.

\bibitem{van96} J. Vandergriff et al., Phys. Lett. B{\bf 365}, 418, 1996.

\bibitem{bar91} H.W. Barz et. al, Nucl. Phys. {\bf 24B}, 211,1991

\bibitem{go89}G.L. Shaw, et al.,
Nature {\bf337},436,1989.


\bibitem{fj85a} E. Farhi and R.L. Jaffe, Phys Rev. D{\bf 30}, 2379, 1984.

\bibitem{be87} M.S. Berger and R.L. Jaffe, Phys. Rev. C {\bf 35}, 213, 1987.

\bibitem{ta88} K. Takahashi and R.N. Boyd, Ap. J. {\bf 327}, 1009, 1988.

\bibitem{fj85} E. Farhi and R.L. Jaffe, Phys Rev. D {\bf 32}, 2452, 1985.

\bibitem{ban83} B. Banerjee et at., Phys. Lett. {\bf B127}, 543, 1983

\bibitem{shu} Y. Shultz et al., Nucl. Phys. {\bf A622}, 404, 1997, and 
references therein. 

\bibitem{ver} P. Vergani et al., Phys. Rev. C{\bf 48},1815, 1993, and 
references therein. 

\bibitem{jpc} R. Jaffe, private communication.

\bibitem{gsref} I.Y. Lee, Proceedings of the workshop on
GAMMASPHERE Physics, World Scientific, 50, 1996,
M. A. Deleplanque, I.Y. Lee and A. O. Macchiavelli, Editors.

\bibitem{ore} Creighton Mine, Shaft number 9,
at the 6800 ft. level. Sudbury, Ontario, Canada.

\bibitem{met} Allende Meteorite, USNM 3529, Smithsonian Institution.

\bibitem{m1} R.V. Morris, et al., Handbook of Lunar
Soils, JSC 19069, PP. 914, Johnson Space Center, Houston, 1983.


\bibitem{trim} TRIM: J.F. Ziegler and
J.P. Biersack, Pergamom Press, NY, 1985.

\bibitem{geref}GEANT: CERN program
Library, W5013, CERN Geneva, Switzerland.

\bibitem{m2} K. Nishiizumi, et al.,
Lunar Planet.
Sci. XXVIII, 1027, 1997.

\bibitem{m3} L.A. Rancitelli, et al.,
Proc. Lunar
Sci. Conf. 5th, 2185, 1974.

\bibitem{m4} J.N. Goswami and D. Lal, Proc. Lunar Sci.
Conf. 5th, 2643, 1974.

\bibitem{bp} E.K. Shirk and P.B. Price, Astrophys. J {\bf 220}, 719, 1978.

\bibitem{p85} N.A. Porter, et al., 
Nature {\bf 316}, 49, 1985.

\end{references}
\end{document}